\documentclass{elsart}
\usepackage{graphics}
\usepackage{epstopdf}
\usepackage{epsfig}
\usepackage{amsmath,amssymb}

\begin{document}

\begin{frontmatter}

\title{Cosmic Microwave Background Fluctuations from Gravitational Waves: An Analytic Approach\thanksref{now}}

\author{Jonathan R. Pritchard\corauthref{cor}}
\corauth[cor]{Corresponding author.}
\ead{jp@tapir.caltech.edu}
\address{California Institute of Technology, Mail Code 130-33, Pasadena, CA
91125}
\author{Marc Kamionkowski}
\ead{kamion@tapir.caltech.edu}
\address{California Institute of Technology, Mail Code 130-33, Pasadena, CA
91125}

\thanks[now]{This work was supported in part by NASA NAG5-11985, and DoE DE-FG03-92ER40701.}


\begin{abstract}
 
We develop an analytic approach to calculation of the temperature and polarisation power spectra of the cosmic microwave background due to inflationary gravitational waves.  This approach complements the more precise numerical results by providing insight into the physical origins of the features in the power spectra.  We explore the use of analytic approximations for the gravitational-wave evolution, making use of the WKB approach to handle the radiation-matter transition.  In the process, we describe scaling relations for the temperature and polarisation power spectra.  We illustrate the dependence of the amplitude, shape, and peak locations on the details of recombination, the gravitational-wave power spectrum, and the cosmological parameters, and explain the origin of the peak locations in the temperature and polarisation power spectra.  The decline in power on small scales in the polarisation power spectra is discussed in terms of phase-damping.  In an appendix we detail numerical techniques for integrating the gravitational-wave evolution in the presence of anisotropic stress from free-streaming neutrinos.

\end{abstract}


\end{frontmatter}

\section{Introduction} 
\label{sec:intro}
The evidence from the cosmic microwave background (CMB) for a flat
Universe \cite{Kamionkowski:1994,BOOMERANG,CBI,WMAP,deBernardis:2000,Miller:1999,Hanany:2000,Halverson:2002,Mason:2003,Benoit:2003,Goldstein:2003,Spergel:2003} and a nearly scale-invariant spectrum of primordial
adiabatic perturbations \cite{BOOMERANG,CBI,WMAP,deBernardis:2000,Miller:1999,Hanany:2000,Halverson:2002,Mason:2003,Benoit:2003,Goldstein:2003,Spergel:2003}, in good agreement
with the predictions of inflation \cite{Guth:1982,Hawking:1982,Linde:1982,Starobinsky:1982,Bardeen:1983,inflation}, motivates
additional observational probes of inflation.  One such probe
is the polarisation signature \cite{Kamionkowski:1997,Kamionkowski:1997b,Zaldarriaga:1996,Seljak:1997} of the stochastic
gravitational-wave background \cite{Abbott:1984,Rubakov:1982,Fabbri:1983,Starobinskii:1985} produced during
inflation, which has now become the target of several
ground-based experiments \cite{bicep,quiet,quad,polarbear}, as well as an
Einstein vision experiment in NASA's science roadmap \cite{BeyondEinstein}.

Large-angle CMB temperature fluctuations from these gravitational
waves (tensor metric perturbations) were first considered in
Refs. \cite{Abbott:1984,Rubakov:1982,Fabbri:1983,Starobinskii:1985} 
and the polarisation was first
considered in Ref. \cite{Polnarev:1985}.  Now, the most precise
predictions for these power spectra come from numerical
calculations \cite{cmbfast}.  Like the power spectra for density
perturbations (scalar metric perturbations), which
exhibit wiggles due to acoustic waves in the primordial
baryon-photon fluid,  the temperature and polarisation power
spectra from gravitational waves exhibit wiggles due to
oscillations of tensor modes as they enter the horizon.  The
wiggles in the density-perturbation power spectra were predicted
originally by Sunyaev and Zeldovich \cite{Sunyaev:1970} and Peebles and Yu
\cite{Peebles:1970}, and explained later elegantly with a
semi-analytic approach in a paper by Hu and Sugiyama
\cite{Hu:1995}.

The goal of this paper is to present an analytic
account of the features in the tensor power spectra.  Such an
approach explains the origin of the features in the temperature
and polarisation power spectra and illustrates the dependence of
these features on the tensor power spectrum, cosmological
parameters, and details of the recombination history.  The
intuition provided by such an approach complements the more
precise results of numerical calculations.  In particular, we
explain here the location of the wiggles in the tensor
temperature and polarisation power spectra, and why the bumps in
the curl component of the polarisation are smoother than those
in the curl-free component.  We also show how the amplitude of
the polarisation depends on the details of the recombination
history.  Our approach is analogous to that for scalar modes
given in Ref. \cite{Hu:1995}.  
We discuss how measurement of the locations of these peaks
can provide an independent probe of cosmological parameters.

The organisation of the paper is as follows.  In Section \ref{sec:exact}, we
write the exact equations for CMB fluctuations from tensor
perturbations.  The exact equations consist of the evolution of
the gravitational waves, the visibility function, the source
function, and projection factors.  Next we develop a qualitative understanding of the physics contained in these relations in Section \ref{sec:tale}.  The remainder of the paper
then investigates individually each ingredient in the exact
calculation.  Section \ref{sec:grav} discusses the evolution of the
gravitational-wave perturbation.  Section \ref{sec:rec} discusses the
effect of the recombination history on the power spectrum.
Section \ref{sec:projection} discusses the projection factors, and Section \ref{sec:source} the
source function.  Finally we comment on the dependence on cosmological parameters and detectability in Section \ref{sec:discussion}.  We include two Appendices that discuss the numerical techniques required to evolve the gravitational-wave amplitude in the presence of neutrino anisotropic stress (Appendix \ref{app:anisotropy}), and the application of the WKB approach to gravitational waves evolving through the matter-radiation transition (Appendix \ref{app:wkb}).

\section{Exact Equations}
\label{sec:exact}

Here we present the exact equations required to evaluate the CMB power spectra from gravitational waves \cite{Kamionkowski:1997,Zaldarriaga:1996,Hu:1997a}.  For simplicity, we will restrict ourselves to the case of a flat FRW universe.  Our emphasis will be on small scale structure and so reionisation and its effects on large scales will not be discussed; see Refs. \cite{Zaldarriaga:1997,Ng:1996} for more details on this topic.

To provide the framework for temperature and polarisation anisotropies, we follow the formalism of Ref. \cite{Zaldarriaga:1996}.  For two other useful introductions into the subject see Refs. \cite{Cabella:2004,Lin:2004}.  The CMB radiation field is characterised by the Stokes parameters $I$,$Q$, and $U$.  The fourth Stokes parameter, $V$, is not generated by Thomson scattering, and while it can be generated after last scattering, the expected amplitudes are small \cite{Cooray:2002} and so can be neglected.

While convenient, the Stokes parameters $Q$ and $U$ describing polarisation suffer from being co-ordinate dependent.  Under a right-handed rotation by an angle $\psi$ in the plane perpendicular to the direction $\hat{n}$ of propagation, $Q$ and $U$ transform according to 
\begin{equation}
Q'=Q\cos 2\psi + U \sin 2\psi,
\end{equation}
\begin{equation}
U'=-Q\sin 2\psi +U \cos 2\psi,
\end{equation}
where $\mathbf{\hat{e}'_1}=\cos \psi\mathbf{\hat{e}_1}+\sin \psi \mathbf{\hat{e}_2}$ and $\mathbf{\hat{e}'_2}=-\sin \psi\mathbf{\hat{e}_1}+\cos \psi \mathbf{\hat{e}_2}$.  These transformation laws motivate the combinations $Q\pm i U$ which have definite spin-2.  We can then expand these quantities in terms of the spin-weighted spherical harmonics which are appropriate to describe the projection of these quantities onto the unit sphere,
\begin{equation}
T(\hat{n})=\sum_{lm}a_{T,lm}\,Y_{lm}(\hat{n}),
\end{equation}
 \begin{equation}
(Q\pm iU)(\hat{n})=\sum_{lm}a_{\pm2,lm}\, {}_{\pm2}Y_{lm}(\hat{n}),
\end{equation} 
where $Y_{lm}(\hat{n})$ and ${}_{\pm2}Y_{lm}(\hat{n})$ are the spin-0 and spin-2 spin-weighted spherical harmonics.
These expressions may be inverted to obtain the spherical-harmonic expansion coefficients,
\begin{equation}
a_{T,lm}=\int d\Omega \, Y_{lm}^*(\hat{n})T(\hat{n}),
\end{equation}
\begin{equation}
a_{\pm2,lm}=\int d\Omega \, {}_{\pm2}Y_{lm}^*(\hat{n})(Q\pm iU)(\hat{n}).
\end{equation}
Rather than work in terms of $a_{\pm2,lm}$ it is advantageous to define two rotationally invariant quantities $E$ and $B$ by the relations
\begin{equation}
a_{E,lm}=-(a_{2,lm}+a_{-2,lm})/2,
\end{equation}
\begin{equation}
a_{B,lm}=i(a_{2,lm}-a_{-2,lm})/2.
\end{equation}
These two quantities are equivalent to the curl and grad modes defined in Ref. \cite{Kamionkowski:1997}.  The $E$ mode is invariant under the parity transformation, while the $B$ mode transforms with odd parity.

From the above $a_{X,lm}$, we can form a series of correlation functions that characterise the statistics of the CMB perturbations.  Of the six possible combinations, the TB and EB cross-correlations will vanish unless parity is somehow violated in the early Universe.  The power spectra are defined as the rotationally-invariant quantities,
\begin{equation}
C_{XX'l}=\frac{1}{2l+1}\sum_{m} \langle a_{X,lm}^* a_{X',lm}\rangle.
\end{equation}

Given a formalism to describe the observed CMB perturbations, it is then necessary to calculate a theoretical description of the perturbations.  The starting point for this is to solve the Boltzmann equation for the radiation transfer of photons.  To proceed, we expand the perturbations in Fourier modes of wavevector $\mathbf{k}$.  A full derivation of the necessary equations is beyond the scope of this paper (for details see Refs. \cite{Zaldarriaga:1996,Polnarev:1985,Crittenden:1993}), so we will summarise the important equations below.

Tensor perturbations are assumed to arise, in similar fashion to scalar perturbations, from quantum fluctuations during inflation.  Although our knowledge of this epoch is speculative, we may describe the perturbations by a primordial power spectrum,
\begin{equation}\label{primordialpower}
P_h(k)=A_T k^{n_T -3},
\end{equation}
where $k$ is the comoving gravitational-wave wavenumber and $A_T$ is an amplitude fixed by the energy density during inflation.
The form of this power spectrum is motivated by inflationary theories which generically predict the tensor spectral index $n_T \approx 0$, a nearly scale-invariant spectrum.  If the process generating the perturbations is Gaussian, then this power spectrum encodes all information about the distribution.

The evolution equations for the tensor modes may be derived from the Einstein equations and are \cite{Bond:1996}
\begin{equation}\label{generaltensoreom}
\ddot{\chi}_{ij}+2\frac{\dot{a}}{a}\dot{\chi}_{ij}+k^2 \chi_{ij}=16\pi G a^2 \pi_{ij}.
\end{equation}
Here, $\chi_{ij}$ is a symmetric, traceless perturbation to the spatial part of the metric, and $\pi_{ij}$ is the tensor part of the anisotropic stress.  Here and throughout, overdots denote derivatives with respect to conformal time, and $a(\tau)$ is the scale factor normalised to unity today. 

The temperature and polarisation anisotropies induced by an equal mixture of tensor modes of $+$ and $\times$ polarisation with amplitude $h$ may be described in terms of the variables $\Delta_X(\tau_0,\mathbf{\hat{n}},\mathbf{k})$, where X=(T,E,B).  These have dependence on both $\phi$ and $\theta$, which motivates the use of new variables $\tilde{\Delta}_T(\tau,\mu,k)$ and $\tilde{\Delta}_P(\tau,\mu,k)$ dependent only on $\theta$, first introduced by Polnarev \cite{Polnarev:1985}.  Here, $\mu=\hat{n}\cdot\hat{k}$ is the angle between the direction $\hat{n}$ of propagation of the photon and the wavevector $\hat{k}$ of the tensor mode.  The relation between these two sets of variables is detailed in Ref. \cite{Zaldarriaga:1997}.  For our purposes it is enough to use the Polnarev variables for calculating sources; the sources are then simply related to the original variables.  

The evolution of a single Fourier mode $k$ satisfies the Boltzmann equations,
\begin{equation}\label{boltzmannT}
\dot{\tilde{\Delta}}_T + ik\mu\tilde{\Delta}_T = -\dot{h}-\dot{\kappa}[\tilde{\Delta}_T - \Psi],
\end{equation}
\begin{equation}\label{boltzmannP}
\dot{\tilde{\Delta}}_P + ik\mu\tilde{\Delta}_P= -\dot{\kappa}[\tilde{\Delta}_P + \Psi],
\end{equation}
\begin{equation}\label{psidefn}
\Psi\equiv\left[\frac{1}{10}\tilde{\Delta}_{T0} + \frac{1}{7}\tilde{\Delta}_{T2} + \frac{3}{70}\tilde{\Delta}_{T4} - \frac{3}{5}\tilde{\Delta}_{P0} + \frac{6}{7}\tilde{\Delta}_{P2} - \frac{3}{70}\tilde{\Delta}_{P4}\right].
\end{equation}
Here, we have defined the differential cross section for Thomson scattering as $\dot{\kappa}=a n_ex_e \sigma_T$, where $n_e$ is the electron number density, $x_e$ is the ionisation fraction, and $\sigma_T$ is the Thomson cross section.  The total optical depth between a conformal time $\tau$ and $\tau_0$ is given by integrating $\dot{\kappa}$ to obtain $\kappa(\tau,\tau_0)=\int_{\tau}^{\tau_0}\dot{\kappa}(\tau)d\tau$.  The multipole moments of temperature and of polarisation are defined by $\Delta(k,\mu)=\sum_l (2l+1)(-i)^l\Delta_l(k)P_l(\mu)$, where $P_l(\mu)$ is the Legendre polynomial of order $l$.  This decomposition converts Eqs. \eqref{boltzmannT} and \eqref{boltzmannP} into an infinite hierarchy of equations connecting higher moments to lower moments.
These equations have solutions \cite{Zaldarriaga:1997}

\begin{equation}\label{lost}
\Delta_{Tl}=\sqrt{\frac{(l+2)!}{(l-2)!}}\int_{0}^{\tau_0} d\tau S_T(k,\tau)\frac{j_l(x)}{x^2},
\end{equation}

\begin{equation}\label{lose}
\Delta_{El}=\int_{0}^{\tau_0} d\tau S_{P}(k,\tau)\left[-j_l(x)+j_l''(x)+\frac{2 j_l(x)}{x^2} +\frac{4 j_l'(x)}{x}\right],
\end{equation}

\begin{equation}\label{losb}
\Delta_{Bl}=\int_{0}^{\tau_0} d\tau S_{P}(k,\tau)\left[2 j_l'(x)+\frac{4 j_l(x)}{x}\right],
\end{equation}
where $j_l(x)$ is the spherical Bessel function.
In these expressions, $x=k(\tau_0-\tau)$.
Defining the visibility function $g(\tau)=\dot{\kappa}e^{-\kappa}$, the sources are given by
\begin{equation}\label{sourceT}
S_T(k,\tau)=-\dot{h}e^{-\kappa} + g \Psi,
\end{equation}
\begin{equation}\label{sourceP}
S_P(k,\tau)=-g\Psi,
\end{equation}
and the power spectra by
\begin{equation}\label{CXXl}
C_{XX'l}=(4\pi)^2\int k^2dk P_h(k)\Delta_{Xl}(k)\Delta_{X'l}(k).
\end{equation}

It is straightforward to show that for statistically equal distributions of left and right circularly polarised gravitational waves, the TB and EB cross-correlations vanish.  If there is a preference for either polarisation, then a non-zero TB and EB correlation will be observed \cite{Lue:1999}.

Equations for the evolution of the ionisation fraction, and hence the optical depth, exist, but will not be dealt with here.  Details of the relevant equations and useful analytic approximations for the optical depth and ionisation fraction may be found in Ref. \cite{Jones:1985}.

Where necessary we assume a fiducial $\Lambda$CDM cosmology with $\Omega_{b}=0.05$, $\Omega_{DM}=0.25$, $\Omega_{\Lambda}=0.7$, and with Hubble parameter parameterised by $h=0.72$.

The above set of equations forms the basis for our problem.  Having written expressions for the power spectra, we must now exploit a mixture of physical and mathematical approximations to bring out their implications.

\section{A Tale of Tensor Modes}
\label{sec:tale}

Let us now try to obtain an intuitive understanding of how the features in the power spectra arise.  This will help motivate the approximations that follow in later sections.  Useful discussions of how polarisation is generated are given in Refs. \cite{Cabella:2004,Dodelson:2003,Hu:1997b}.

\begin{figure}[htbp]
\begin{center}
\includegraphics[scale=0.4]{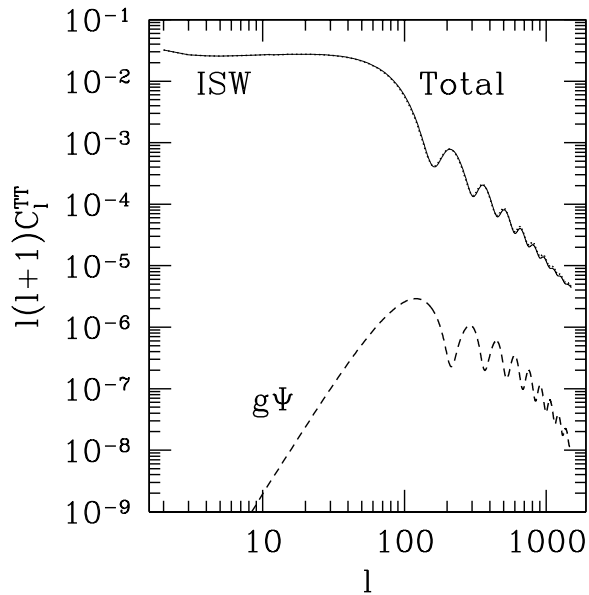}
\caption{Comparison of power generated by the two source terms for temperature anisotropy.  Plotted are the total power (solid line), ISW term only (dotted line), and $g\Psi$ term only (dashed line).  The $g\Psi$ term is essentially negligible at all $l$.  The normalisation here, and in all plots, is specified by setting $A_T=1/8(4\pi)^2$ and $n_T=0$.}
\label{fig:sourcecomp}
\end{center}
\end{figure}

First let us discuss the temperature power spectrum.  The temperature multipole moments due to an individual gravitational-wave of wavenumber $k$ observed at a conformal time $\tau_0$ are 
\begin{equation}\label{lost2}
\Delta_{Tl}=\sqrt{\frac{(l+2)!}{(l-2)!}}\int_{0}^{\tau_0} d\tau \left(-\dot{h}e^{-\kappa} + g \Psi\right)\frac{j_l(x)}{x^2},
\end{equation}
with $x=k(\tau_0-\tau)$.  The second of the sourcing terms is localised to the surface of last scattering (SLS) by the visibility function; as a consequence of the restricted range this term is small and may be neglected at all angular scales.  Between $l=200$ and $l=800$, the contribution from the $g\Psi$ term falls off more slowly than the integrated Sachs-Wolfe (ISW) term allowing it to become marginally relevant.  At lower and higher $l$, the power generated by the second term dies off rapidly and is totally negligible (Fig. \ref{fig:sourcecomp}). The first term, which dominates this integral, involves an integral from the SLS to the present day.  Its form tells us that the temperature power spectrum is sensitive to the evolution of gravitational waves from the SLS to today and not to the recombination history.  This term is a form of integrated Sachs-Wolfe (ISW) effect that describes the anisotropy generated by changing gravitational potentials.  
We can understand this effect by recalling that a gravitational-wave alternately stretches and compresses space as it oscillates.  A photon travelling past the gravitational-wave loses energy when its wavelength is stretched, but gains energy when its wavelength is reduced.  If the gravitational-wave amplitude evolves over the course of the oscillation, the photon will undergo a net change in energy.  Tensor modes decrease steadily in amplitude and oscillate after horizon entry.  As such, a photon travelling along the crest of a phase front will slowly gain energy as the overall amplitude of the gravitational-wave decreases.  Photons travelling at an angle to the mode
experience further red and blue shifting as they propagate through different phase regions.  Their energy oscillates as a consequence.  Between the SLS and today, the period-averaged amplitude of the tensor mode decreases, and so the mean energy of the photon increases.    

If we consider a late time, so that the amplitude of the tensor modes is essentially zero, then we see that the final energy of the photon is determined by whether it started its journey from the SLS at a trough or crest in the tensor mode.  Photons starting at a crest will have gained more energy and appear hotter than average and vice versa for those starting at a trough.  This simplistic picture is modified by the effect of power free-streaming from one angular scale to another as the Universe expands, which tends to smooth the resultant power spectra.  

The situation is very different for the polarisation anisotropies.  These are generated by expressions of the form
\begin{equation}\label{losx}
\Delta_{Xl}=\int_{0}^{\tau_0} d\tau\left(-g\Psi \right)P_{Xl}[k(\tau_0-\tau)].
\end{equation}
Here, the source is very firmly localised to the SLS and so is sensitive to the thermal history and gravitational-wave evolution at that time.  This is sensible. Treating the early radiation bath as unpolarised (as we expect from suppression of anisotropy during the tightly-coupled regime), then polarisation is generated by Thomson scattering of an anisotropic intensity distribution.  Where does this anisotropy come from?  In the rest frame of the scattering electron, photons arrive from all directions from a mean distance determined by the mean free path of photons at recombination.  In propagating, these photons experience the ISW effect, discussed in the case of the temperature spectrum, and so arrive at the scatterer with altered temperatures.  The resulting anisotropic temperature distribution is scattered, generating polarisation which free-streams to the present epoch.

\begin{figure}[htbp]
\begin{center}
\includegraphics[scale=0.5]{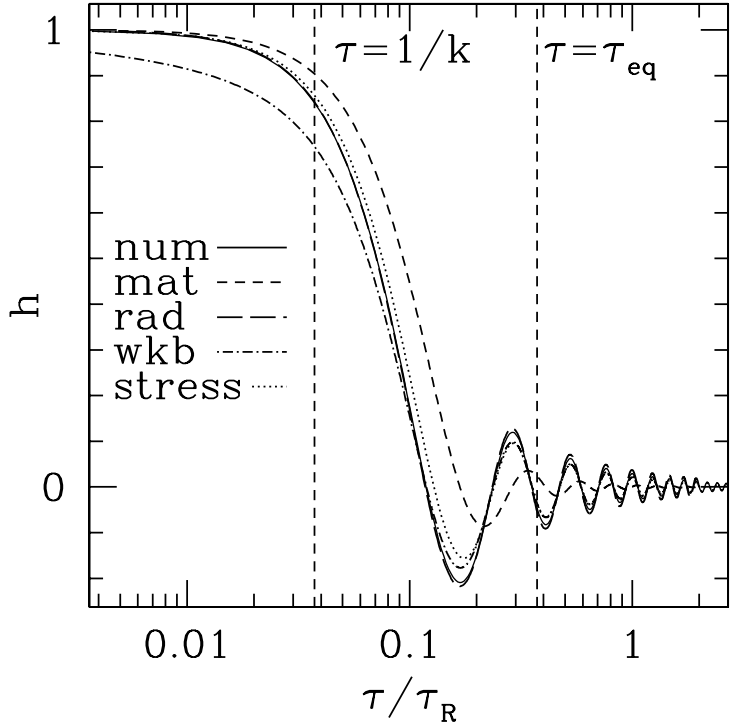}
\caption{Evolution of a gravitational-wave.  Wavenumber $k$ satisfies $k\tau_{\rm{eq}}=10$.  Shown solutions are numerical without anisotropic stress (solid curve), numerical with anisotropic stress (dotted curve), radiation (long dashed curve), matter (short dashed curve), and WKB (dot-dashed curve).  The two vertical lines denote $\tau=1/k$ and $\tau_{\rm{eq}}$. }
\label{fig:hplot}
\end{center}
\end{figure}

In this way, we can understand the power spectrum.  For modes with wavelengths much larger than the horizon size at last scattering, incident photons experience very little ISW before the last scattering event and little polarisation is generated.  Optimal ISW and thus maximal polarisation is generated by modes that enter the horizon at the time of penultimate scattering.  The amplitude of the gravitational-wave decays most rapidly immediately on horizon entry (see Fig. \ref{fig:hplot}) before settling into oscillation with a slowly decreasing amplitude.  Modes that enter the horizon before penultimate scattering lead to photons whose ISW samples this slowly decaying regime.  Hence, they generate significant polarisation, but less than for the optimal case.  Note that the time between penultimate and last scattering will be about the width of the surface of last scattering.

Translating this into the form of the polarisation power spectrum, we expect a slow increase in power at large scales peaking at the scale of the horizon at penultimate scattering.  Immediately after this, we expect a large drop in polarisation corresponding to the transition between modes that enter the horizon between penultimate and last scattering and those that do not.  Next, we expect a steady decline in power as modes have entered the horizon before penultimate scattering and so redshifted away before the ISW effect is generated.  This region will show a transition in slope between modes that entered the horizon in the matter- and radiation-dominated epochs.  On scales smaller than the mean free path at recombination, the power will drop sharply as phase cancellation between differing crests and troughs becomes important.

\begin{figure}[htbp]
\begin{center}
\includegraphics[scale=0.5]{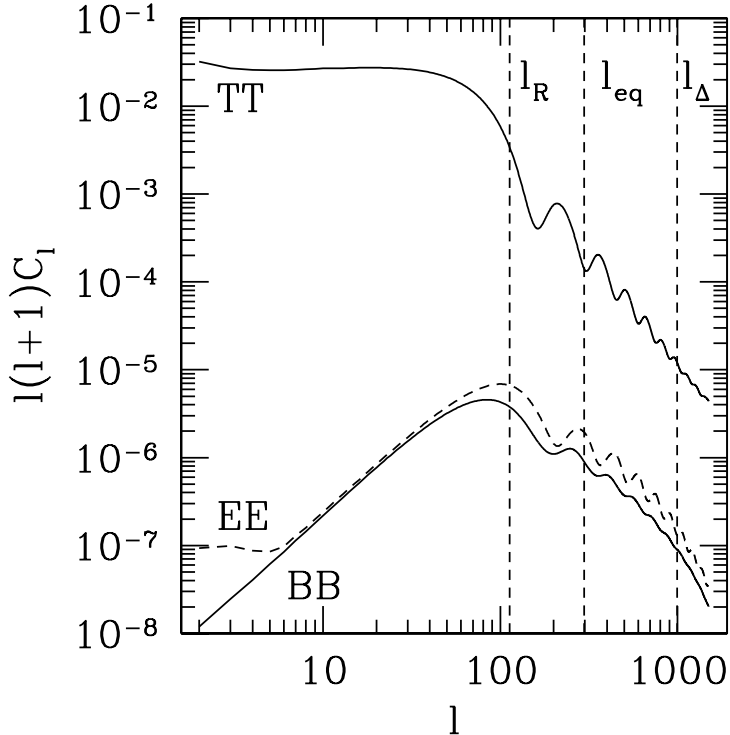}
\caption{Tensor power spectra.  Curves from top to bottom are $C^{TT}_l$, $C^{EE}_l$, and $C^{BB}_l$.   Vertical lines indicate important angular scales, from left to right: horizon at recombination, $\tau_R$, horizon at matter-radiation equality, $\tau_{eq}$, and the width of the last-scattering surface, $\Delta\tau_R$.}
\label{fig:tensor_ref}
\end{center}
\end{figure}

In this discussion, it is important to realise that only three scales have entered the problem.  These are the comoving horizon at recombination, the horizon at matter-radiation equality, and the width of the last-scattering surface.  Fig. \ref{fig:tensor_ref} shows how the features in the power spectrum correspond to these scales.

\section{Gravitational-wave Evolution}
\label{sec:grav}

Expansion of the Universe leads to damping of the tensor modes as described by the term proportional to $\dot{h}$ in Eq. \eqref{generaltensoreom}.  This is the usual redshifting of radiation.  In addition, the tensor modes may be sourced by anisotropic stress, $\pi_{ij}$.  It has been shown \cite{Bond:1996,Weinberg:2003} that anisotropic stress generated by free-streaming neutrinos acts to provide viscosity, further damping the tensor modes.  This effect is important only while the energy density in neutrinos is a significant fraction of the total energy; i.e., during the radiation-dominated epoch. 

The tensor modes may be decomposed into two independent polarisation states, $h_\times$ and $h_+$.  With this decomposition and a source term appropriate for neutrino anisotropic stress, we have
\begin{equation}\label{heom_stress}
\ddot{h}_i+2\frac{\dot{a}}{a}\dot{h}_i+k^2 h_i=-24 f_\nu(\tau)\left(\frac{\dot{a}(\tau)}{a(\tau)}\right)^2 \int^\tau_0 K[k(\tau-\tau')]\dot{h}_i(\tau') d\tau',
\end{equation}
where $i=+,\times$, and $f_\nu\equiv\bar{\rho}_\nu/\bar{\rho}$ with $\bar{\rho}$ the unperturbed density, and $K(s)$ is given by
\begin{equation}
K(s)\equiv -\frac{\sin s}{s^3} -\frac{3 \cos s}{s^4}+\frac{3 \sin s}{s^5}.
\end{equation}

To a first approximation, we may neglect the effect of anisotropic stress, though it should be included in detailed calculations.  Without the source term, analytic solutions for Eq. \eqref{heom_stress} in pure radiation and matter cosmologies may be expressed in terms of the spherical Bessel function $j_l(x)$,
\begin{equation}\label{hrad}
h_{\rm{rad}}(\tau)=h(0)j_0(k\tau)=h(0)\frac{\sin k\tau}{k\tau},
\end{equation}
\begin{equation}\label{hmat}
h_{\rm{mat}}(\tau)=3h(0)\frac{j_1(k\tau)}{k\tau}.
\end{equation} 
In a mixed radiation and matter dominated universe, the solution follows $h_{\rm{rad}}$ initially before asymptotically becoming similar to $h_{\rm{mat}}$.  The initial radiation dominated phase introduces a phase shift into $h_{\rm{mat}}$ as now the boundary conditions do not preclude the spherical Neumann solution to the unsourced Eq. \eqref{heom_stress}.  When calculating the power spectra it is important to get this phase, which determines the peak positions, correct.  This point was understood but not implemented in Ref. \cite{Turner:1993} and included implicitly by others \cite{Wang:1996}.

The behaviour of these solutions is shown in Fig. \ref{fig:hplot} and splits into three main regimes.  When $k\tau \ll1$, $h$ evolves slowly and is approximately constant.  Once $k\tau\approx 1$, the amplitude decays away rapidly before entering an oscillatory phase with slowly decreasing amplitude, when $k\tau \gg1$.  Physically, this corresponds to a mode that is frozen beyond the horizon until its wavelength is of order the horizon size at which point it enters the horizon and redshifts rapidly with the expansion of the Universe.

Recombination occurs shortly after the Universe becomes matter dominated.  For modes that enter the horizon during the matter-dominated regime and so have evolved little in the radiation-dominated epoch, we expect $h_{\rm{mat}}$ to be a good description.  For modes that entered during radiation domination, we expect that the transition from radiation to matter domination will affect the evolution significantly. 

The matter-radiation transition can be accounted for in a variety of ways.  Most simple is to assume that the transition is instantaneous and to match the amplitude and derivative of $h$ on the boundary. This will be a good approximation for waves with wavelength much longer than the time taken for the transition to take place.

\begin{align}
h_{\rm{instant}}=\left\{
\begin{array}{ll}
 j_0(k\tau), &\: \tau<\tau_{\rm{eq}}, \\
 (\tau_{\rm{eq}}/\tau)[A j_1(k\tau)+B y_1(k\tau)], &\: \tau>\tau_{\rm{eq}},
\end{array}\right.
  \end{align}
with 
\begin{equation}
A=\frac{\frac{3}{2}k\tau_{\rm{eq}}-\frac{1}{2}k\tau_{\rm{eq}}\cos(2k\tau_{\rm{eq}})+\sin(2k\tau_{\rm{eq}})}{k^2\tau_{\rm{eq}}^2},
\end{equation}
\begin{equation}
B=\frac{2-2k^2\tau_{\rm{eq}}^2-2\cos(2k\tau_{\rm{eq}})-k\tau_{\rm{eq}}\sin(2k\tau_{\rm{eq}})}{2k^2\tau_{\rm{eq}}^2}.
\end{equation}
Alternatively, we may consider the situation where the wavelength of the gravitational-wave is much shorter than the transition time.  In this case, the gravitational-wave sees the background expansion vary slowly and a WKB approach is appropriate.  Ng and Speliotopoulis \cite{Ng:1995} first presented this approach, although they were primarily interested in late time asymptotic limits and so neglected the behaviour near the classical turning point.  Here we generalise their result making use of the uniform Langer solution for the WKB problem \cite{Bender:1978}.  The result is
\begin{multline}
h(\tau)=\frac{\Gamma(k\tau)^{-1/4}}{\tau^{1/2}(\tau+2)}\left(\frac{3}{2}S_0(\tau)\right)^{1/6} \\ \times\left\{2\sqrt{\pi}C_2 \rm{Ai}\left[\left(\frac{3}{2}S_0(\tau)\right)^{2/3}\right]+\sqrt{\pi}C_1 \rm{Bi}\left[\left(\frac{3}{2}S_0(\tau)\right)^{2/3}\right] \right\},
\end{multline}
with
\begin{equation}
\Gamma(s)=\frac{1}{4}+\frac{2s}{s+2k}-s^2,
\end{equation}
and
\begin{equation}
S_0(\tau)=\int_{k\tau}^{k\tau_T} \sqrt{\Gamma(s)}\, \frac{ds}{s}.
\end{equation}
Here, $\tau_T$ is the solution to $\Gamma(k\tau)=0$, Ai and Bi are Airy functions, and $C_1$ and $C_2$ are constant coefficients set by the boundary conditions $h(0)=1$ and $\dot{h}(0)=0$.  For technical reasons, these boundary conditions must be extrapolated to small $\tau$ via asymptotic approximation to Eq. \eqref{heom_stress} and then applied.  Care must be taken in evaluating the above expressions when $\tau>\tau_T$.  These details are discussed further in Appendix \ref{app:wkb}.  This WKB expression reproduces the phase of $h$ in both radiation and matter dominated regimes, but underestimates the amplitude.  The close agreement between the WKB and anisotropic-stress curves in Fig. \ref{fig:hplot} is a numerical coincidence.
   
Other approaches exist to handle this transition from radiation to matter in a more pragmatic fashion \cite{Turner:1993,Wang:1996}.

We can get the scaling of $h$ from a simple argument.  Before horizon entry, the amplitude $h$ of a gravitational-wave is constant.  After horizon entry, the gravitational-wave redshifts with the expansion as radiation and scales as $h\sim1/a$.  Hence, the amplitudes of a gravitational-wave today and at horizon entry are related by $h_{\rm{today}}/h_{\rm{entry}}=a_{\rm{entry}}/a_{\rm{today}}$.  Taking $h_{\rm{entry}}$ to be independent of $k$, we have $h_{\rm{today}}\propto a_{\rm{entry}}$.  Horizon entry occurs when $a_{\rm{entry}}H_{\rm{entry}}=k$, and so from the scaling of $H$ in the matter- and radiation-dominated epochs we obtain $a_{\rm{entry}}\propto k^{-1}$ when radiation dominated and $a_{\rm{entry}}\propto k^{-2}$ when matter dominated.  Thus, we obtain the scalings,
\begin{align}\label{hscale}
h \propto\left\{
\begin{array}{ll}
 1, &\: k <1/\tau_0, \\
  k^{-2}, &\:  1/\tau_{\rm{eq}} > k >1/\tau_0, \\
  k^{-1}, &\: k>1/\tau_{\rm{eq}}.
\end{array}\right.
  \end{align}
 This result agrees with both the instantaneous-transition and WKB solutions when $\tau\gg\tau_{\rm{eq}}$.  These scaling relations form the basis for scaling of the power spectrum.  We expect $l(l+1)C^{TT}_l$ to scale as \cite{Starobinskii:1985,Turner:1993}
\begin{align}\label{cltscale}
l(l+1)C^{TT}_l \propto\left\{
\begin{array}{ll}
 1, &\: l <l_R, \\
  l^{-4}, &\:  l_{\rm{eq}} > l >l_R, \\
  l^{-2}, &\: k>l_{\rm{eq}} .
\end{array}\right.
\end{align}
It has been claimed \cite{Starobinskii:1985,Atrio-Barandela:1994} that there should be an extra region, $l>l_{\Delta}$, in which the width of the last-scattering surface becomes important and due to phase-damping the scaling goes as $l^{-6}$ .  However the dominant source of temperature anisotropy is the ISW effect, which is insensitive to the recombination history, and so we do not expect to see this behaviour in the temperature power spectrum.  One way to see this is to examine the kernel in Eq. \eqref{lost}.  On small scales, which enter the horizon before $\tau_R$, the finite rise time of $e^{-\kappa}$ alters the weight in the integral by a nearly constant factor.  For all other modes, it is sufficient to simply truncate the range of the integral to between $\tau_R$ and $\tau_0$, effectively imposing instantaneous recombination. On the other hand, the polarisation anisotropy is generated near the SLS and will show phase cancellation dependent on the width of the SLS.  We will return to this point later.

In the absence of reionisation, at low $l$ the power spectra for the polarisation grow as $l^2$ \cite{Hu:1997a}.  From this and the above scaling arguments, we would expect the power spectrum to scale as
\begin{align}\label{clxscale}
l(l+1)C^{XX}_l \propto\left\{
\begin{array}{ll}
 l^2, &\: l <l_R, \\
  l^{-2}, &\:  l_{\rm{eq}} > l >l_R, \\
  1, &\: l_{\rm{eq}}<l<l_\Delta,\\
  l^{-4}, &\: l>l_\Delta,
\end{array}\right.
\end{align}
with X=(E,B).
The effect of phase-damping extends to much lower $l$ than would be indicated by these simple dimensional arguments.  Consequently, the region of constant power $l_{\rm{eq}}<l<l_\Delta$ is never visible in calculated spectra, but is lost in the transition to the phase-damping regime.
 
\begin{figure}[htbp]
\begin{center}
\includegraphics[scale=0.5]{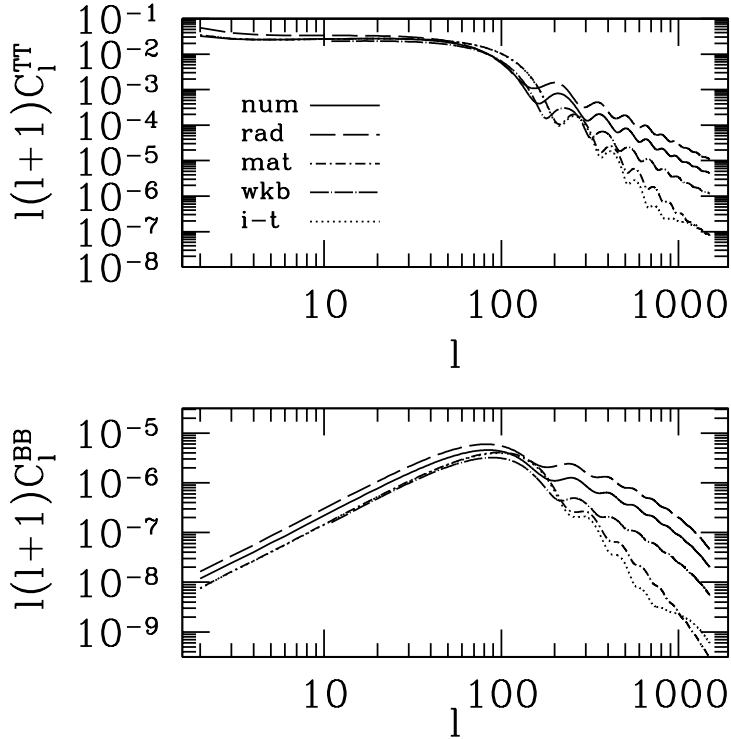}
\caption{T and B power spectra calculated using approximate forms for the gravitational-wave amplitude $h$.  Plotted are the results using $h$ from the full numerical calculation (solid curve) and from the radiation-dominated (long dashed curve), matter-dominated (dot-short dashed curve), instantaneous-transition (dotted curve), and the WKB (dot-long dashed curve) approximations. }
\label{fig:gwavespec}
\end{center}
\end{figure}

The above expressions for the gravitational-wave amplitude are used to generate the power spectra displayed in Fig. \ref{fig:gwavespec}.  The plots are normalised by taking $n_T=0$ and setting $A_T=1/8(4\pi)^2$.  All of the plots show the same scaling relation at low $l$.  This regime is dominated by modes that have not entered the horizon at recombination and so are approximately constant.  At low $l$ $h_{\rm{mat}}$ underestimates the power, while $h_{\rm{rad}}$ overestimates the power.  This is a consequence of the contribution of the modes that have entered the horizon that are evolving in a mixed radiation-matter universe and so have amplitudes intermediate to the predictions of these two approximations.

Moving above the peak at $l_R$, we clearly see the different scaling relations between the matter and radiation approximations.  Recall that at these high $l$s, we expect the main contribution to come from modes that entered the horizon in the radiation-dominated epoch, and so $h_{\rm{rad}}$ should be a good approximation.  The WKB result shows a transition between following the matter-dominated curve to behaving more like the radiation-dominated form, though with reduced amplitude.  This reduction in amplitude is an unfortunate characteristic of the WKB solution and is not significant to understanding the physics.  The WKB solution serves as a nice bridge between matter- and radiation-dominated epochs.  The instantaneous solution fails to be useful on scales with wavelength short compared to the transition time-scale.

These curves display the scaling expected from Eq. \eqref{hscale}, but we see that in the numerical case, excluding anisotropic stress, we never observe the full $l^{-4}$ scaling for a matter-dominated regime.  The combination of recombination occurring soon after matter-radiation equality and the Universe becoming matter dominated only slowly means that the power spectra damp more slowly, closer to $l^{-3}$, for scales $1/\tau_R<k<1/\tau_{\rm{eq}}$.  The presence of matter also causes peak positions to shift to smaller scales over the fully radiation-dominated case indicative of the phase shift that the transition introduces in $h$.

For the purposes of reproducing the exact tensor-mode power spectra, we must worry about preserving both the amplitude and phase of the gravitational waves.  The importance of the amplitude is clear in estimating the power correctly.  The phase determines the positions of the maxima and minima in the high-$l$ region of the spectra.  Maxima correspond to gravitational waves whose amplitude was at a maximum or minimum at the SLS; minima correspond to gravitational waves whose amplitude was close to zero at the SLS.  Altering the phase of the gravitational waves shifts the $k$ values for which these maxima occur at the SLS and so shift the features in the CMB. If we wish to understand these features in detail, then we must understand how the phase of the gravitational waves varies with $k$ and how this is mapped onto the power spectrum.  This mapping is the subject of Section \ref{sec:projection}.

\section{Recombination History}
\label{sec:rec}

While the Universe is young and hot, baryons are ionised and tightly coupled to  photons via Thomson scattering.  Once the temperature falls below a few eV, it becomes favourable for electrons and ions to recombine to form neutral molecules.  As the number of charged particles falls, the mean free path of any given photon increases.  Eventually, the mean free path becomes comparable to the horizon size and the photon and baryon fluids are essentially decoupled.  It is at this point in the Universe's evolution that the CMB photons last scatter.

The visibility function describes the probability that a given CMB photon last scattered from a particular time.  In terms of the optical depth $\kappa$, this visibility function is given by
\begin{equation} \label{eq:visibility}
g(\tau)=\dot{\kappa}e^{-\kappa}.
\end{equation}
Numerical calculations show that $g(\tau)$ is sharply peaked during recombination.  This property suggests we approximate the visibility function by a narrow Gaussian for analytic simplicity.  For example,
\begin{equation}\label{eq:gaussianvis}
g(\tau)=g(\tau_R) e^{-\frac{(\tau-\tau_R)^2}{2 \Delta\tau_R^2}},
\end{equation}
determines the visibility function in terms of the conformal time $\tau_R$ of recombination, its width $\Delta\tau_R$, and the amplitude $g(\tau_R)$ at recombination.    

Approximating the visibility by a Gaussian leads to a simple form for the optical depth in the region close to $\tau_R$.  If we write $\kappa$ in the general form $\kappa = \exp[-f(\tau)]$, then consistency with
Eqs. \eqref{eq:visibility} and \eqref{eq:gaussianvis} requires that $\kappa\approx \exp[-(\tau-\tau_R)/\Delta\tau_R]$ and $g(\tau_R)\approx1/(e \Delta\tau_R)$ in the region close to $\tau_R$.  This latter result is essentially a statement about the normalisation of the Gaussian and preserves the total weight of the visibility function for different widths.  Away from recombination, the evolution of the optical depth is a complicated function of the thermal history and not easily approximated.

\begin{figure}[htbp]
\begin{center}
\includegraphics[scale=0.4]{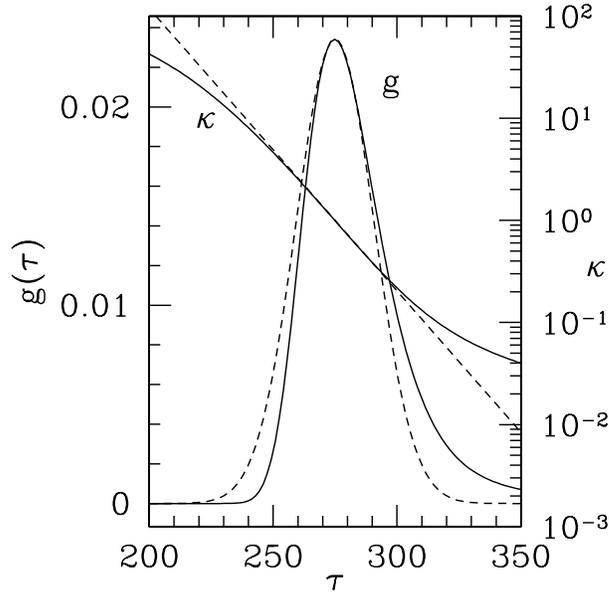}
\caption{Recombination history.  Plotted are the visibility function $g(\tau)$ and the optical depth $\kappa$ calculated numerically (solid curves) and the approximations described in the text (dashed curves) using $\Delta\tau_R=15.7$.}
\label{fig:recomb}
\end{center}
\end{figure}
These approximations for $g(\tau)$ and $\kappa$ are plotted in Fig. \ref{fig:recomb} for the fiducial cosmology with $\Omega_b=0.05$, $\Omega_{\rm{DM}}=0.25$, and $\Omega_\Lambda=0.7$.  For this cosmology, we have $\tau_0=13515\, \rm{Mpc}$, $\tau_R/\tau_0=0.0203$, $\tau_{\rm{eq}}/\tau_0=0.0076$, and $\Delta\tau_R/\tau_0=0.0012$.  While the Gaussian form does a reasonable job of approximating the shape of the peak, the visibility function is clearly skewed and possesses a significant tail.  The combination of these features means the Gaussian approximation will underestimate the power and shift features to slightly smaller angles than in the true power spectrum.

\begin{figure}[htbp]
\begin{center}
\includegraphics[scale=0.5]{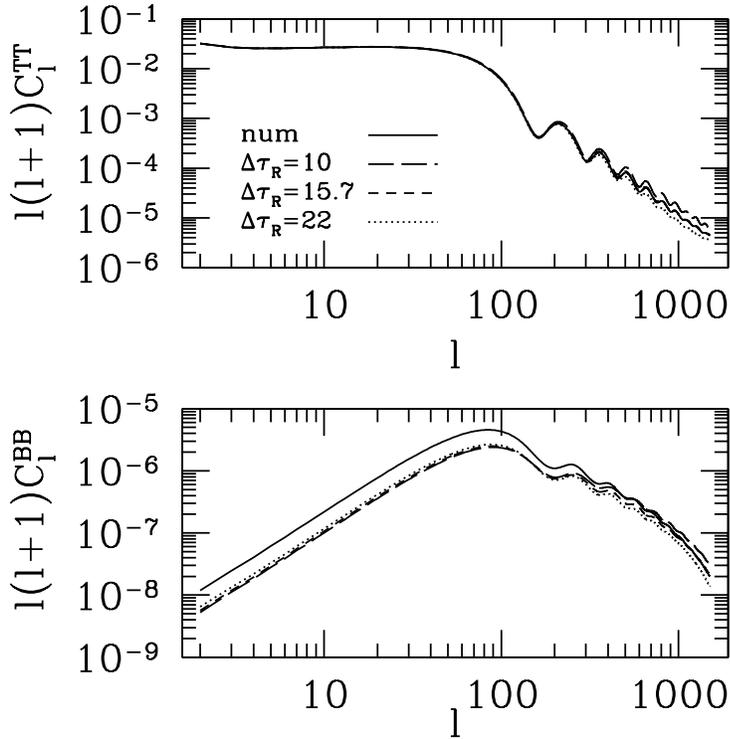}
\caption{Evaluation of the Gaussian approximation for the visibility function.  Three Gaussian derived power spectra are shown for $\Delta\tau_R=10$ (long dashed curve), $15.7$ (short dashed curve), and $22$ (dotted curve).}
\label{fig:gaussiancmb}
\end{center}
\end{figure}

Fig. \ref{fig:gaussiancmb} shows a series of power spectra calculated using the Gaussian approximation.  In each, the correct thermal history is used to calculate the evolution of the source function $\Psi$ with the Gaussian approximation applied when calculating the $\Delta_X$ from Eqs. \eqref{sourceT} and \eqref{sourceP}.  Although not strictly self-consistent, this  isolates the modification of the source due to a changed thermal history from the effect of the visibility function on generating anisotropies.  Source evolution will be considered in Section \ref{sec:source}.

The temperature power spectrum shows no variation with $\Delta\tau_R$ at $l<200$.  Power on these scales is generated via the integrated Sachs-Wolfe effect by modes that only evolve significantly between $\tau_R$ and $\tau_0$ and so are insensitive to the thermal history.  At smaller scales, the modes of interest are evolving over recombination and so contain information about the thermal history.  Modifying the width of the visibility function affects the power spectrum via the $e^{-\kappa}$ term in Eq. \eqref{lost2} which acts to cut the integral off below $\tau_R$.  Widening the SLS makes this cutoff slower which, owing to the concave nature of  $e^{-\kappa}$, leads to less weight in the integral.  This leads to the differences observed in the top panel of Fig. \ref{fig:gaussiancmb}.  This is not phase-damping, and does not alter the scaling of the power spectrum significantly.  In addition to this overall shift in power, larger $\Delta\tau_R$ acts to wash out the bumps and wiggles.  For modes that oscillate rapidly over this rise time, the ISW samples an averaged starting value of $h$ and so is less sensitive to the presence of peaks and troughs.  Power is still generated from the net decrease in the maximum amplitude of $h$ from $\tau_R$ to $\tau_0$, so this effect does not cause a large drop in power as suggested in Refs. \cite{Starobinskii:1985,Atrio-Barandela:1994}.

Fig. \ref{fig:gaussiancmb} shows that the Gaussian approximation leads to a lower polarisation power spectrum.  This difference is a consequence of the long tail to the visibility function, which is not reproduced in the Gaussian approximation.  In these plots, varying $\Delta\tau_R$ does not affect the overall amplitude of the power spectrum.  This is an artefact of using the same $\Psi$ for each plot.  In reality, the amplitude of the polarisation power spectra depends sensitively on $\Delta\tau_R$, as will be shown in Section \ref{sec:source}.  These plots show that varying $\Delta\tau_R$ in the Gaussian approximation does not affect the shape at low $l$, but a wider width leads to a sharper fall off in power at high $l$.  This is a feature of phase-damping, which will be discussed in Section \ref{sec:source}.

None of the three values used precisely reproduces the decline of the true power spectrum, which is seen to fall off more rapidly than the approximations.  This seems to be a consequence of the tail to the visibility function.  In keeping with expectation, the peaks in the high-$l$ region are found at slightly higher $l$ in the approximations than the numerical result.

\section{Projection Factors}
\label{sec:projection}

The power spectra that we observe today are projections of the temperature and polarisation anisotropies at the last-scattering surface.  By inspection of Eqs. \eqref{lost}, \eqref{lose}, and \eqref{losb} we can define three projection terms,
\begin{equation}\label{projectT}
P_{Tl}(x)=\frac{j_l(x)}{x^2},
\end{equation}
\begin{equation}\label{projectE}
P_{El}(x)=-j_l(x)+j_l''(x)+\frac{2 j_l(x)}{x^2} +\frac{4 j_l'(x)}{x},
\end{equation}
\begin{equation}\label{projectB}
P_{Bl}(x)=2 j_l'(x)+\frac{4 j_l(x)}{x}.
\end{equation}
Typically the argument of these terms is $k(\tau_0-\tau)$, the look-back time scaled by the wavenumber, reflecting that these are projections from the point of origin onto today's sky. 

\begin{figure}[htbp]
\begin{center}
\includegraphics[scale=0.4]{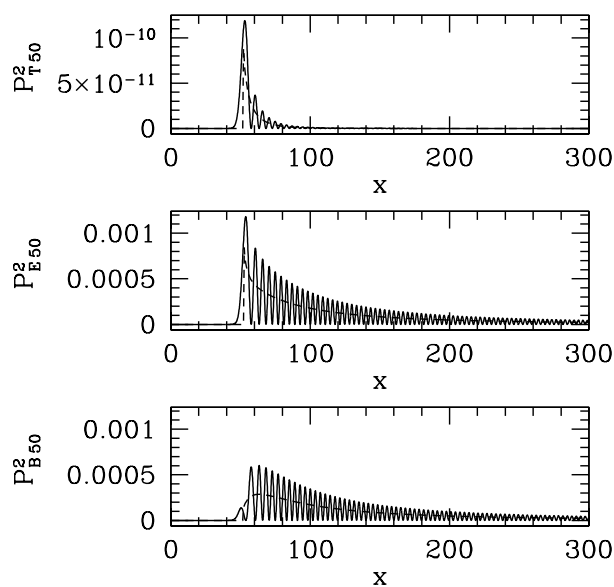}
\caption{Projection terms for $l=50$.  Dashed curve shows approximations.}
\label{fig:projection}
\end{center}
\end{figure}

\begin{figure}[htbp]
\begin{center}
\includegraphics[scale=0.4]{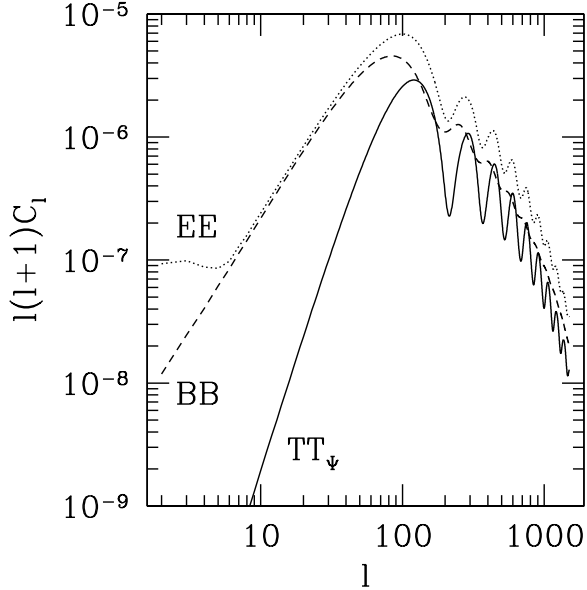}
\caption{Power spectra generated solely due to the source term $g\Psi$.  Temperature (solid curve), E mode (dotted curve), and B mode (dashed curve).}
\label{fig:polcomp}
\end{center}
\end{figure}

The different forms of the projection factors, plotted in Fig. \ref{fig:projection}, help explain many of the features seen in the power spectra (Fig. \ref{fig:polcomp}).

First consider the E projection factor as a simple example.  A sharp peak occurs at $x\approx l$. This tells us that the value of $C^{TT}_l$ at $l=50$ is determined by the behaviour of the source function at $x\approx l$.  The polarisation source function is strongly peaked around $\tau=\tau_R$, which implies the behaviour at last scattering of the mode with wavenumber $k\approx l/(\tau_0-\tau_R)$ dominates the contribution to $C_l$.  If the projection factor was a Dirac delta function, this would be the whole story.  However, the projection factor has a significant tail for $x>l$ signifying that modes with larger wavenumber also contribute power to this angular scale.

From this, we can see that the sharper the spike at $x\approx l$, the sharper the features seen in the power spectrum.  A wider peak mixes in modes of different phases blurring the spectra.  Noting that the B projection factor lacks a sharp peak, we expect the B power spectra to contain blurred features relative to the E spectra.  In addition, as its maximum is at higher $x$, we expect features in the source to be shifted to smaller $l$ than in the E spectrum.  This sort of argument has some validity with the T spectrum, but is there complicated by the extended nature of the source term.

The complicated form of the projection factors makes analytic progress difficult.  Similarly, their oscillatory behaviour makes numerical integration very slow at high $k$ values.  CMBFAST implements a scheme for fast numerical integration.  Here we discuss time averaging the projection functions to get a useful analytic envelope.

Widely known approximations for the spherical Bessel functions in the cases $x\gg l$ and $x\ll l$ exist and are in common usage.  For our purposes, though, we are most interested in the case where $x\approx l$; i.e., in the vicinity of the peaks of the projection factor.  A relatively simple approximate form may be derived which is valid in the regime $x>l$ \cite{Debye:1909},
\begin{equation}\label{besselapprox}
j_l(x)=\frac{1}{\sqrt{x^2\sin\alpha}}\cos\left[x(\sin\alpha-\alpha\cos\alpha)-\pi/4\right],
\end{equation}
where $\cos\alpha=(l+1/2)/x$.  This can be shown to reduce to the usual $j_l(x)\approx\sin(x-l\pi/2)/x$ for $x\gg l$.  Approximations valid in the regime $x\approx l$ exist, but are more complicated and will not significantly improve on this level of approximation.

This approximation is still complicated and shows strong oscillation.  When we calculate power spectra we will be interested in quantities of the form $\left[P_X(x)\right]^2$.  To proceed, we substitute Eq. \eqref{besselapprox} into the projection factor and then average the squared projection function over a full cycle to extract the variation of the envelope.  Time averaging makes use of the relations $\langle\sin^2x\rangle=\langle\cos^2x\rangle=1/2$ and $\langle\sin x \cos x\rangle=0$.  This envelope may then be further simplified by assuming $x\gg1$ and $l\gg1$.  The resulting envelope functions are not pretty, but may be numerically integrated by a standard routine.  They are
\begin{equation}\label{approxprojectT}
\langle P_T(x)^2\rangle\approx\frac{1}{2 x^5 \sqrt{x^2-l^2}},
\end{equation}
 \begin{multline}\label{approxprojectE}
\langle P_E(x)^2\rangle\approx\big\{-16(l+l^2-x^2)(l+12 l^3+8 l^4+8 x^4-4(x+2 l x)^2)^2\\
 +(-16 l^5(2+l)-4(1+2 l)(3+10 l)x^4 +32 x^6 \\ 
+(-1+8 l(1+2l))(x+2 l x)^2)^2\big\}/\big\{512x^5(-l(1+l)+x^2)^{9/2}\big\},
\end{multline}
 \begin{equation}\label{approxprojectB}
\langle P_B(x)^2\rangle\approx \frac{(12 l^2-8 x^2)^2 -16(l+2 l^2-2 x^2)^2(l+l^2-x^2)}{x^3(-4 l^2+4 x^2)^{5/2}}.
\end{equation}
 The form for $\langle P_T(x)^2\rangle$ is consistent with that quoted in Ref. \cite{Zaldarriaga:1995}.

One slight complication is that these approximations show divergent behaviour as $x\rightarrow l$ making it necessary to arbitrarily restrict the domain to $x>l+a$, where $a$ is an arbitrary cutoff of order unity. Fig. \ref{fig:projection} shows these approximations and the cutoff.  This cutoff procedure behaves best in the case of the B projection factor which is already decreasing as $x\rightarrow l$.  The E and T projection factors have considerable weight near to the peak, and so care must be taken in selecting the cutoff.

To apply these projection factors, we approximate the anisotropy term,
\begin{equation}
\Delta_{Xl}=\int^{\tau_0}_{0} d\tau\, g(\tau) \Psi(\tau) P_{Xl}[k(\tau_0-\tau)],
\end{equation}
by the expression,
\begin{equation}
\Delta_{Xl}\approx P_{Xl}[k(\tau_0-\tau_R)]\int^{\tau_0}_{0} d\tau\, g(\tau) \Psi(\tau),
\end{equation}
where we have pulled the projection factor out from the integral.  This should be a good approximation provided that the projection factor varies slowly relative to the source term.

\begin{figure}[htbp]
\begin{center}
\resizebox{8cm}{!}{\includegraphics{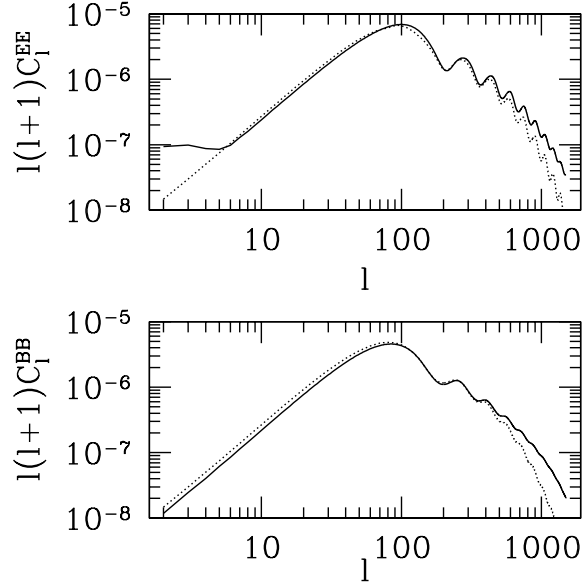}}
\caption{Comparison of power spectra with exact (solid curve) and approximate (dashed curve) projection factors.}
\label{fig: projectionspectra}
\end{center}
\end{figure}
The power spectra calculated under this approximation are shown in Fig. \ref{fig: projectionspectra}. Note that two approximations are combined here.  The projection term has been pulled from the integral and then the time averaged expression used in place of the exact form.  The inadequacy of this approximation at large $l$ can be seen.  This is to be expected, as in this regime the relevant $k$ modes oscillate significantly over the width of the last-scattering surface.  At values $l<500$, the approximation is appropriate. 

A significant discrepancy in the EE power spectrum is visible at $l<6$.  In making the approximations to produce Eqs. \eqref{approxprojectT}, \eqref{approxprojectE}, and \eqref{approxprojectB}, we have discarded information about very large angular scales.  The rise in power in the EE power spectrum is sensitive to these large angular scales \cite{Zaldarriaga:1997} and so is not reproduced by our approximations.

\section{Source Evolution}
\label{sec:source}

Having discussed the recombination history and the mechanics of projection, we turn to the core issue of how the source itself evolves.  

Those modes that enter the horizon close to recombination evolve only slowly over the width of the last-scattering surface.  For these modes, it is possible to derive analytic approximations to the source functions that occur in the expressions for $\tilde{\Delta}_X$.  Here, we follow the approach of Zaldarriaga and Harari \cite{Zaldarriaga:1995}, but see also Ref. \cite{Keating:1998}.  Preceding recombination, the optical depth $\kappa$ is large and the photons are tightly coupled to the baryonic fluid.  In this regime, we may expand the Boltzmann equations for the temperature and polarisation multipoles in powers of $\dot{\kappa}^{-1}$.  Keeping terms to first order in $\dot{\kappa}^{-1}$, we obtain the equations,
\begin{equation}
\dot{\tilde{\Delta}}_{T0}  = -\dot{h}-\dot{\kappa}[\tilde{\Delta}_{T0}-\Psi], 
\end{equation}
\begin{equation}
\dot{\tilde{\Delta}}_{P0}  = -\dot{\kappa}[\tilde{\Delta}_{P0}+\Psi],
\end{equation}
\begin{equation}
\dot{\tilde{\Delta}}_{Tl} =0\;, \: l\geq 1,
\end{equation}
\begin{equation}
\dot{\tilde{\Delta}}_{Pl} =0\;, \: l\geq 1.
\end{equation}

Using these equations together with the definition of $\Psi$ gives us an expression for the time evolution of the source function within this tightly-coupled limit,
\begin{equation}
\dot{\Psi}+\frac{3}{10}\dot{\kappa}\Psi = -\frac{\dot{h}}{10}.
\end{equation}
Thus, an approximate solution for $\Psi$ is
\begin{equation}
\Psi(\tau)=\int^{\tau_0}_0 d\tau' \,\left(-\frac{\dot{h}(\tau')}{10}\exp\left[-\frac{3}{10} \bigg(\kappa(\tau')-\kappa(\tau)\bigg)\right]\right).
\end{equation}
If we assume that the visibility function is approximately Gaussian during recombination, then we may approximate $\dot{\kappa}\approx -\kappa/\Delta\tau_R$.  This allows a change of variable to $x=\kappa(\tau')/\kappa(\tau)$ which leads to
\begin{equation} \label{eq:approxPsi}
\Psi(\tau)=-\frac{\dot{h}(\tau_R)}{10}e^{\frac{3}{10}\kappa(\tau)}\Delta\tau_R\int^\infty_1 \frac{dx}{x}e^{-\frac{3}{10}\kappa x}.
\end{equation}
In taking the gravitational-wave driving term outside of the integral, we have assumed that $h$ varies slowly over the visibility function.  This approximation is only valid for $k\ll1/\Delta\tau_R$.  At larger wavenumbers, the rapid oscillation of $\Psi$ over the visibility function makes this a poor approximation.  An improvement is to replace $\dot{h}(\tau_R)$ in Eq. \eqref{eq:approxPsi} with its value averaged over the visibility function
\begin{equation}\label{averageh}
\langle\dot{h}(\tau)\rangle=\int^{\tau_0}_{0} d\tau\,g(\tau)\dot{h}(\tau)\approx \dot{h}(\tau_R)e^{-(k\Delta\tau_R)^2/2}.
\end{equation}
In calculating the right-hand side, we have treated $\dot{h}$ as an oscillatory function with a slowly-varying envelope.  Integrating an oscillatory function over a Gaussian leads to the function evaluated at the Gaussian's peak multiplied by a decaying exponential.  This exponential decay has a clear physical interpretation.  For modes with $k>1/\Delta\tau_R$, the source function oscillates rapidly across the visibility function.  Hence, different regions in the SLS contribute to the observed polarisation with different phases leading to cancellation and a decrease in the observed power.  We will refer to this cancellation as phase-damping.  While present in the scalar modes, this effect is overwhelmed by diffusion damping \cite{Hu:1996}.  Diffusion damping makes the effective visibility function scale dependent and always sufficiently narrow that phase-damping is not important.  For the tensor modes, phase-damping provides the dominant process for damping on small scales. 

\begin{figure}[htbp]
\begin{center}
\includegraphics[scale=0.6]{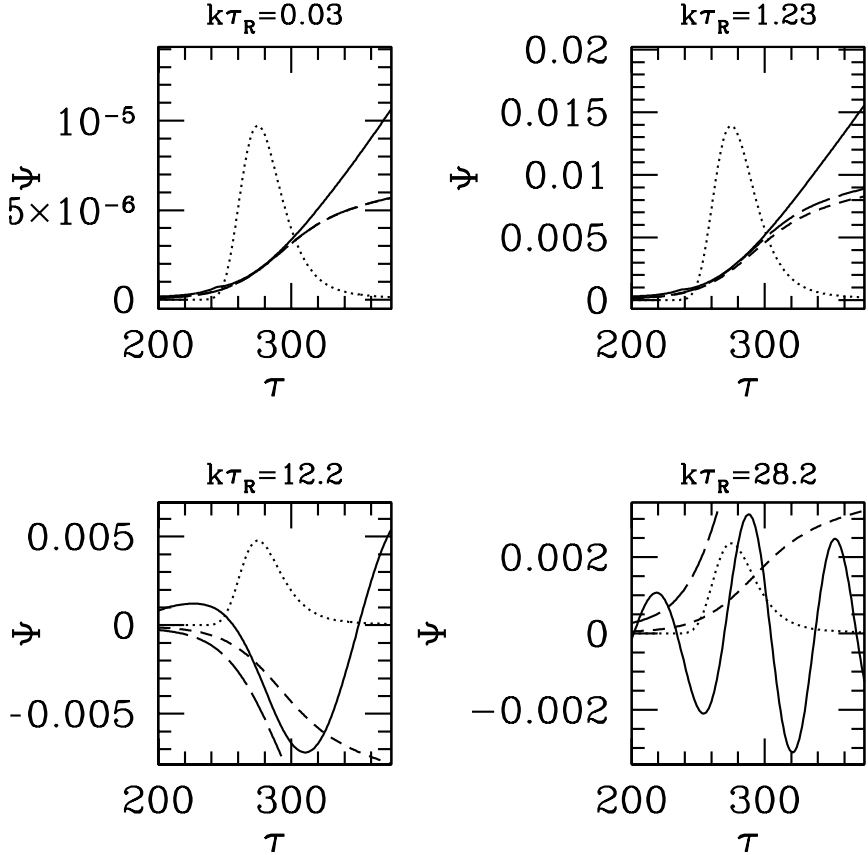}
\caption{Comparison of Eq. \eqref{eq:approxPsi} (long dashed curve), Eq. \eqref{eq:approxPsi} with damping term (short dashed curve), and numerical calculation of $\Psi$ (solid curve).  Four different values of $k$ are plotted:  $k\tau_R=0.03$, $1.23$, $12.2$, and $28.2$.  An arbitrarily scaled visibility function (dotted curve) has been plotted in each panel to guide the eye.}
\label{fig:sourceplot}
\end{center}
\end{figure}
Fig. \ref{fig:sourceplot} shows the behaviour of $\Psi$ and our analytic approximations for four values of $k$.  For small $k$, the approximation closely mirrors the growth of $\Psi$ in the region where the visibility function has weight.  At larger $k$, the source function is seen to oscillate across the width of $g(\tau)$; this is not reproduced by either approximation.  This should not be cause for concern as we now discuss. 

\begin{figure}[htbp]
\begin{center}
\includegraphics[scale=0.4]{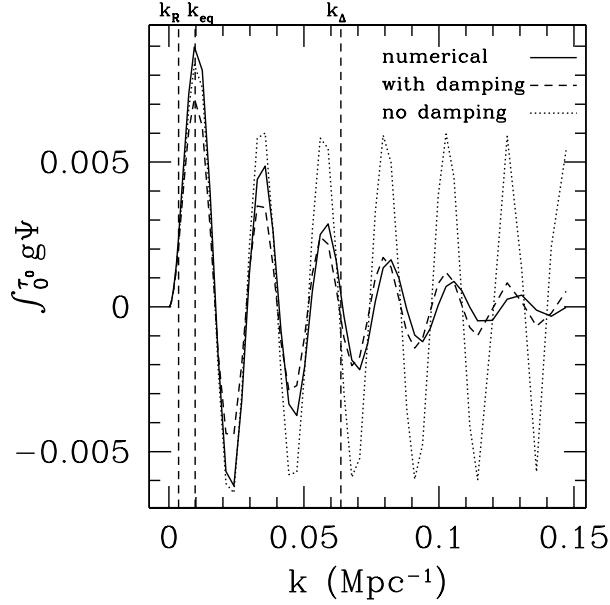}
\caption{Comparison of $\int^{\tau_0}_0 g\Psi$ for the numerical $\Psi$ (solid curve), the approximation for $\Psi$ without phase-damping (dotted curve) and the approximation for $\Psi$ with phase-damping (dashed curve).  Vertical lines indicate (from left to right) $1/\tau_R$, $1/\tau_{\rm{eq}}$, and $1/\Delta\tau_R$.}
\label{fig:psicomp}
\end{center}
\end{figure}
The quantity of real interest is the anisotropy that this source generates.  This is calculated from Eqs. \eqref{lose} and \eqref{losb}.  For the given polarisation, X=(E,B), we have
\begin{equation}
\Delta_{Xl}(k)=\int^{\tau_0}_{0} d\tau\, g(\tau) \Psi(\tau) P_{Xl}[k(\tau_0-\tau)].
\end{equation}
First we pull the projection term outside of the integral assuming that it varies slowly over the width of the visibility function,
\begin{equation}\label{eq:anisotropyX}
\Delta_{Xl}(k)\approx P_{Xl}[k(\tau_0-\tau_R)]\int^{\tau_0}_{0} d\tau\, g(\tau) \Psi(\tau).
\end{equation}
Note that $\Psi$ appears only through an integral over the visibility function.  Provided that our approximation can reproduce this integrated behaviour, the fact that it fails to reproduce the temporal oscillation is unimportant.  Fig. \ref{fig:psicomp} indicates the close agreement between the integrated source function and our approximation, provided that phase-damping is taken into account.  

Having checked the validity of our approximation we substitute for $\Psi$ in \eqref{eq:anisotropyX} using Eqs. \eqref{eq:approxPsi} and \eqref{averageh} giving
\begin{equation}
\Delta_{Xl}(k)= P_{Xl}[k(\tau_0-\tau_R)] \frac{1}{10}\dot{h}(\tau_R)\Delta\tau_Re^{-(k\Delta\tau_R)^2/2}\int^\infty_0 d\kappa\, e^{-\frac{7}{10}\kappa}  \int^\infty_1 \frac{dx}{x}e^{-\frac{3}{10}\kappa x}.
\end{equation}
The integrals evaluate to $(10/7) \log(10/7)$, which leads to the final result
\begin{equation}\label{analyticanisolow}
\Delta_{Xl}= P_{Xl}[k(\tau_0-\tau_R)]\dot{h}(\tau_R)\Delta\tau_Re^{-(k\Delta\tau_R)^2/2}\left(\frac{1}{7}\log\frac{10}{3}\right).
\end{equation}
This result is proportional to the width $\Delta\tau_R$ of recombination as might be expected.  During recombination, photons will travel for a distance of order $\Delta\tau_R$ before scattering.  This is the time available for the quadrupole which sources the polarisation to grow, and so we expect a result proportional to $k\Delta\tau_R$.

Extending this result to calculate the power spectrum is straightforward.  We have Eq. \eqref{CXXl},
\begin{equation}
C_{Xl}=(4\pi)^2\int k^2dk P_h(k)\left[\Delta_{Xl}(k)\right]^2.
\end{equation}
Applying our expression for $\Delta_{Xl}(k)$ yields the final result for this Section,
\begin{equation}\label{analyticlowcl}
C_{Xl}=(4\pi)^2\left(\frac{1}{7}\log\frac{10}{3}\right)^2  \int k^2dk P_h(k) P_{Xl}[k(\tau_0-\tau_R)]^2\dot{h}(\tau_R)^2\Delta\tau_R^2e^{-(k\Delta\tau_R)^2}.
\end{equation} 

\begin{figure}[htbp]
\begin{center}
\includegraphics[scale=0.5]{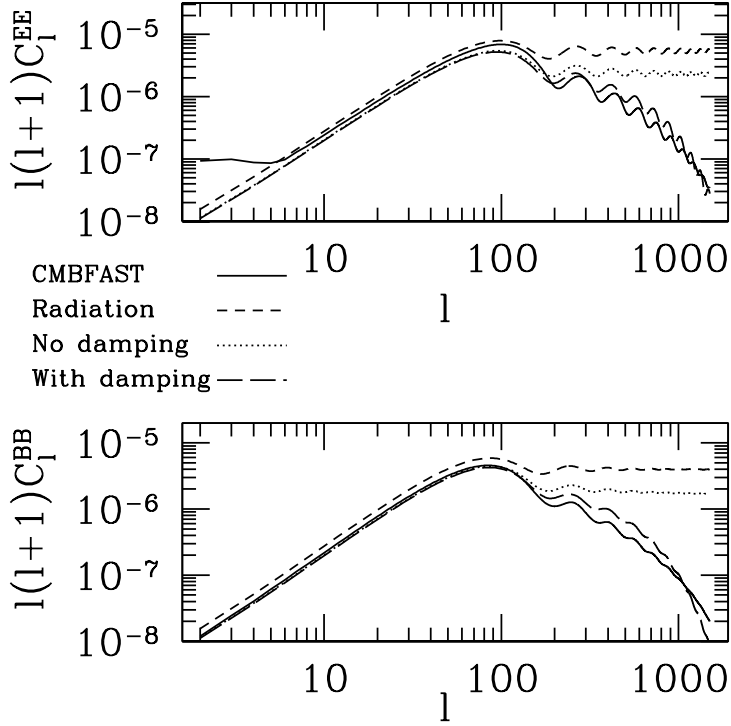}
\caption{Comparison of E and B power spectra for CMBFAST (solid curve) and analytic approximations:  $h$ for radiation epoch (short dashed curve), $h$ calculated numerically (dotted curve), and $h$ calculated numerically with exponential damping (long dashed curve).}
\label{fig:cllow_num}
\end{center}
\end{figure}
Fig. \ref{fig:cllow_num} compares the result of CMBFAST with that from Eq. \eqref{analyticlowcl}.  All plots have been calculated for the fiducial cosmology with $\Omega_{b}=0.05$, $\Omega_{DM}=0.25$, and $\Omega_{\Lambda}=0.7$.  Values for $\tau_R$ and $\Delta\tau_R$ were chosen by fitting by eye to the visibility function produced by CMBFAST.  Radiation and numerical forms for $h$ without phase-damping are plotted alongside a numerical form for $h$ with phase-damping and the results of the full numerical calculation.

Agreement between all of the solutions is good at low $l$ where our assumptions are most valid and the anisotropy is building slowly.  Similarly, the position of the main peak is accurately reproduced, although it becomes clear that the E projection-factor approximation is less reliable than the B-mode one.  Beyond the main peak, the effects of the different forms for $h$ become apparent.   

Without phase-damping, the radiation-dominated form of $h$ leads to an almost flat power spectrum, consistent with Eq. \eqref{clxscale}, with a pattern of bumps and wiggles roughly in phase with those in the full numerical calculation.  Moving to the numerical form of $h$ leads to some damping over the radiation case.  This reflects the increased redshifting which occurs when matter becomes important between $\tau_{\rm{eq}}$ and $\tau_R$.  We never observe the expected matter-dominated scaling of $l^{-2}$, although the presence of matter does lead to a slight decline in power as $l$ increases.  For this cosmology, the ratio $\rho_r/\rho_m$ evaluated at  $\tau_R$ is 0.29 showing that the radiation content is still significant at recombination.

Neither of these forms reproduces the rapid decline in power at large $l$, which is not surprising as we have yet to include phase-damping.  Once this is included, the shape of the power spectrum is much closer to that of the numerical calculation showing a sharp decline in power above $l\approx300$ and reproducing the position of the peaks to reasonable accuracy.  Power in the range $l=150$ to $l=600$ is slightly overestimated.  For $l>600$, the limitations of our projection-factor approximations become apparent with the very rapid drop in power previously observed in Section \ref{sec:projection}.

\section{Discussion}
\label{sec:discussion}

Here, we discuss the information that the features in the tensor power spectrum contain and how detection would complement our existing understanding of the early Universe.  In contrast to the scalar modes, the tensor modes contain very clean information about the evolution of the Universe.  The features of the scalar spectra are a result of the oscillation of the matter-radiation fluid during the period up to recombination.  The scalar spectra encode information about the sound speed of the baryon-radiation fluid, the baryon fraction, and other information about the particle content of the Universe \cite{Dodelson:2003,Hu:2001}.  In contrast, the features of the tensor spectrum are determined solely by the wave motion of the evolving gravitational waves.  They primarily contain information about the expansion rate during the early Universe.  Through their overall angular scale, both spectra encode basic information about the epoch and duration of recombination and the geometry of the Universe.

The first peak of the tensor polarisation spectrum, occurring at $l\approx 90$ for the B modes and $l\approx 105$ for the E modes, is determined by the horizon scale at recombination.  The exact angular scale is determined by this along with the redshift of recombination and the geometry of the Universe.  This information can be determined from the scalar modes allowing a direct measurement of the horizon scale.  The amplitude of the tensor power spectra is directly related to the energy scale of inflation.  Slow-roll inflation, parameterised by the energy scale $E_i$ of inflation, predicts a B-mode power spectrum with a peak at $l\approx90$ and peak amplitude \cite{Hu:2003}
\begin{equation}
\Delta B_{\rm{peak}}=0.024\left(\frac{E_i}{10^{16}\rm{GeV}}\right)^2\mu \rm{K}.
\end{equation}
Measuring this main peak is the subject of several experimental endeavours \cite{bicep,quiet,polarbear} with hope of detection in the not-so-distant future.  

For all models consistent with WMAP constraints to the energy scale of inflation, features after the main peak are sub-dominant to the lensed B-mode signal \cite{Hu:2003,Wang:2002}.  This necessitates the use of algorithms to clean the polarisation maps and recover the tensor signal.  Techniques using maximum likelihood \cite{Seljak:1999,Hirata:2003a,Hirata:2003b} and quadratic estimators \cite{Hu:2002,Kesden:2003} have been advanced to deal with this problem.  Even so, this will complicate precision measurements of the tensor B mode after the main peak.
    
Measuring the overall amplitude after the main peak should recover the scaling relations discussed in Section \ref{sec:grav}.  The breaks in the different regimes yield the horizon scales of matter-radiation equality and the width of the SLS.  This in itself is enough information to constrain a cosmological model yielding $\Omega_m$ and $\Omega_r$.  Neutrino anisotropic stress further damps power on small scales and will make detection more difficult while adding extra information about the neutrino fraction.

The positions of the peaks and troughs on small scales contain information about the phase of the gravitational-wave at recombination.  This in turn depends upon the early expansion rate.  
The acoustic peaks in the scalar power spectrum can be used as a standard ruler; the wiggles in the tensor power spectrum can have the same utility, but operating on a different range and spacing.  Measuring these wiggles would better constrain our cosmology, though it is doubtful that scales small enough to probe the very early Universe will be observed in the foreseeable future.

\section{Conclusions}
\label{sec:conclusion}

This investigation has probed the individual elements that compose the calculation of the tensor power spectra.  Using a variety of approximations, we have obtained a semi-analytic expression which qualitatively reproduces the behaviour of more detailed calculation.  While the approximation is clearly not suitable for precise comparison with data, it serves to illustrate the important physics in an intuitive fashion.   We have shown that the features of the power spectrum may be explained with reference to three main scales: the horizon size at recombination, the horizon size at matter-radiation equality, and the width of the SLS.  The first two scales determine the evolution of the tensor modes; the latter relates to the effect of the thermal history on the generation of anisotropies.  The shapes of the polarisation spectra show most sensitivity to the width of the SLS through the phenomenon of phase-damping that dominates on small scales.  The effect of the thermal history on the temperature spectra is much less dramatic, affecting the amplitude and smoothing on small scales.  We have seen that the position of the peaks and troughs in the power spectrum relate to the phase of the gravitational-wave at recombination.  It is of interest that we do not see modes displaying matter-dominated behaviour in our calculations.  This would allow the tensor spectra to probe the radiation content at recombination.
Useful scaling relations have been developed and clarified.  We hope this paper will aid in general understanding of the tensor modes and inspire future experimental efforts.


\appendix
\section{Numerical Evolution of Gravitational Waves with Anisotropic Stress}
\label{app:anisotropy}

In this Appendix, we return to the question of anisotropic stress.  In the early Universe, free-streaming neutrinos provide the main source of anisotropic stress.  After recombination, photons free stream and can also contribute, though the energy density in radiation is falling fast and the effect is negligible.  Working from Eq. \eqref{generaltensoreom} and standard expressions for the energy density of a distribution of relativistic massless particles, an integro-differential equation describing the evolution of the tensor modes may be derived \cite{Weinberg:2003},
\begin{equation}\label{heom_stress2}
\ddot{h}+2\frac{\dot{a}}{a}\dot{h}+k^2 h=-24 f_\nu(\tau)\left(\frac{\dot{a}(\tau)}{a(\tau)}\right)^2\int^\tau_0 K[k(\tau-\tau')]\dot{h}(\tau') d\tau',
\end{equation}
where $f_\nu\equiv\bar{\rho}_\nu/\bar{\rho}$ with $\bar{\rho}$ the unperturbed density, and $K(s)$ is given by
\begin{equation}
K(s)\equiv -\frac{\sin s}{s^3} -\frac{3 \cos s}{s^4}+\frac{3 \sin s}{s^5}.
\end{equation}
The new term acts to damp the amplitude of $h$ and can be seen to have the form of a convolution over the mode's past history of the kernel $K(s)$ and the ``velocity" $\dot{h}$.  The linear dependence on $f_\nu$ means that the damping term will become negligible in the matter-dominated regime where $f_\nu \propto a^{-1}$.  In the radiation-dominated epoch, though, the neutrino and total energy densities scale in the same way leading to $f_\nu=0.40523$.  This suggests that the damping term will primarily affect those modes that enter the horizon well within the radiation-dominated epoch.  In consequence, it will affect the power spectrum only at high $l$ where these modes are the dominant contributors. 

The right-hand side is also damped by the $(\dot{a}/a)^2$ term which scales as $\tau^{-2}$ in both the matter- and radiation-dominated epochs.  
 
Numerical integration of Eq. \eqref{heom_stress2} is possible after recasting the integro-differential equation as a set of coupled Volterra integral equations \cite{Brunner:1988}.  These coupled equations may then be integrated using standard techniques \cite{numericalrecipes}.

Given an integro-differential equation of the form,
\begin{equation}\label{intdiffeqn}
y^{(r)}(t)=f\big(t,y(t),\dots,y^{(r-1)}(t)\big) +\int^t_0 K\big(t,s,y(s),\dots,y^{(r)}(s)\big)ds,
\end{equation}
and defining $z_k(t)=y^{(k)}(t)$ for $k=0,\dots,r-1$, we may recast Eq. \eqref{intdiffeqn} as the set of first-order Volterra integral equations,
\begin{equation}
z_r(t)=\int^t_0 K\left(t,s,y(s),\dots,y^{(r-1)}(s)\right)ds,
\end{equation}
\begin{equation}
z_{r-1}(t)=y^{(r-1)}_0 + \int^t_0 \big(f(s,z_0(s),\dots,z_{r-1}(s)) + z_r(s)\big)ds,
\end{equation}
\begin{equation}
z_k(t)=y^{(k)}_0+\int^t_0 z_{k+1}(s) ds,
\end{equation}
where the $y^k_0$ are the relevant initial conditions.

Making the co-ordinate transformation $t=k\tau$ in Eq. \eqref{heom_stress2}, we can apply this formalism to obtain
\begin{equation}
z_2(t)=\int^t_0 K_n(t-s) z_1(s) ds,
\end{equation}
\begin{equation}
z_1(t)=y^1_0 + \int^t_0 \left(-2\frac{a'(s)}{a(s)} z_1(s) -z_0(s)) + z_2(s)\right) ds,
\end{equation}
\begin{equation}
z_0(t)=y_0+\int^t_0 z_1(s) ds,
\end{equation}
where primes denote differentiation with respect to $t$, and we have defined $z_1=h'$, $z_0=h$, and \begin{equation}
K_n(s)=-24 f_\nu(t)\bigg(\frac{\dot{a}(t)}{a(t)}\bigg)^2 K(t-s).
\end{equation}

Fig. \ref{fig:hplot} shows the results of numerical integration of Eq. \eqref{heom_stress2}.  The inclusion of anisotropic stress damps the wave during the radiation-dominated regime leading to decreased amplitude and a slightly shifted phase. 

Of primary interest here is the effect of the anisotropic stress by the time of recombination.  It is traditional to calculate a transfer function relating the amplitude and phase of the numerical solution to that of the matter-dominated solution $h_{\rm{mat}}(\tau)=3j_1(k\tau)/k\tau$.  At recombination, radiation is still important and this analytic form is a relatively poor approximation.  For illustrative purposes, we will numerically calculate the amplitude ratio $A$ and phase shift $\Psi$ between numerical calculations of $h$ with and without the effects of anisotropic stress evaluated at the time of recombination.  To calculate $A$ at a given $\tau$, we first calculate $\Psi$ and then numerically fit $h_{\rm{stress}}(k\tau)$ with $A\,h_{\rm{no-stress}}(k\tau+\Psi)$ over the period containing $\tau$.  This avoids the oscillation that results if we seek to obtain the amplitude ratio by simply dividing $h_{\rm{stress}}(k\tau)$ by $h_{\rm{no-stress}}(k\tau)$. 

\begin{figure}[htbp]
\begin{center}
\includegraphics[scale=0.4]{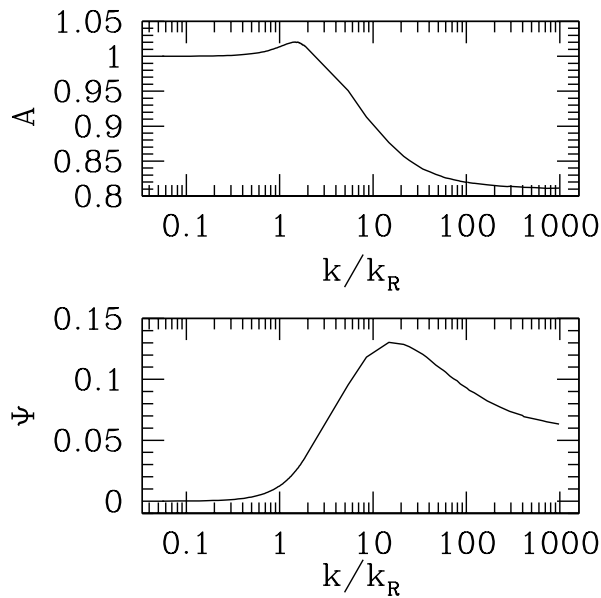}
\caption{Transfer function for gravitational waves when anisotropic stress is present.  Top panel shows the variation of the amplitude ratio A between the case with anisotropic stress and without anisotropic stress.  Bottom panel shows how the phase shift $\Psi$ between the two cases varies with k.}
\label{fig:transferplot}
\end{center}
\end{figure}

The results of this calculation (Fig. \ref{fig:transferplot}) illustrate that anisotropic stress introduces a $k$-dependent damping asymptoting to a factor of $A\sim0.81$.  The phase shift introduced remains small and reaches a maximum value of $\Psi\sim0.13\, \rm{rad}$.
The amplitude ratio is unity while the gravitational-wave remains within the horizon.  Around horizon entry its rises slightly above unity before decreasing asymptotically to $A\sim0.81$.  This slight rise is a consequence of the anisotropic stress which, like viscosity acting on a pendulum, slows the initial decrease of the gravitational-wave and leads to a lower final amplitude of oscillation.  The phase difference between damped and undamped cases grows after horizon entry as a consequence of the slower evolution of the damped wave.  It peaks and begins to asymptote to a constant value for modes that entered the horizon sufficiently before matter-radiation equality to reach their asymptotically damped form.

\begin{figure}[htbp]
\begin{center}
\includegraphics[scale=0.5]{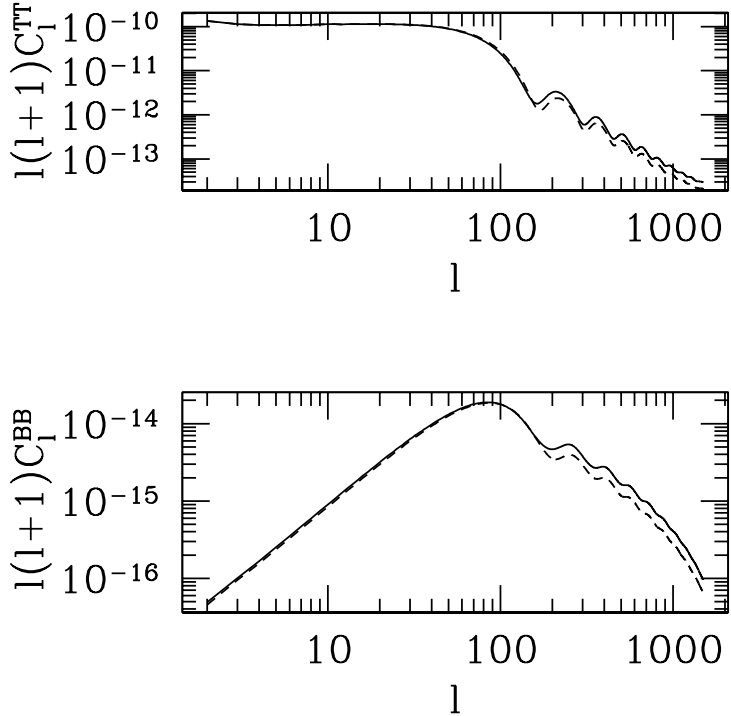}
\caption{T and B power spectra incorporating anisotropic stress (dashed curve) and without anisotropic stress (solid curve).  Damping in the power is clearly seen on small scales. }
\label{fig:cmb_stress}
\end{center}
\end{figure}

CMBFAST may be modified to incorporate the evolution equation \eqref{heom_stress2} and the resulting power spectra calculated.  Fig. \ref{fig:cmb_stress} shows the effects for the T and B power spectra.  The spectra are essentially unchanged at low $l$.  Only at high $l$ do we see a suppression of power from the additional damping.  Damping only occurs in waves that have evolved significantly during the radiation-dominated epoch, so this makes sense.


\section{WKB Solution}
\label{app:wkb}

If the scale factor changes more slowly than the evolution of the gravitational-wave, then WKB techniques become a sensible method of approximation.  In this Appendix, we detail the application of this approach to the evolution of a gravitational-wave mode through the matter-radiation transition.  The WKB approach was first applied in this context by Ng and Speliotopoulos \cite{Ng:1995}, although the solution presented here is of a slightly more general nature.

We begin with the equation of motion for $h(\tau)$ and specific initial conditions that we wish to solve.  Working with the dimensionless variables $\eta=(\sqrt{2}-1)\tau/\tau_{\rm{eq}}$ and $q=k\tau_{\rm{eq}}/(\sqrt{2}-1)$, we have 
\begin{equation}\label{heom_ap}
\ddot{h}+2\frac{\dot{a}}{a}\dot{h}+q^2h=0,
\end{equation}
\begin{equation}
h(0)=1,
\end{equation}
\begin{equation}
\dot{h}(0)=0,
\end{equation}
with overdots indicating differentiation with respect to $\eta$.
The behaviour of the scale factor in a universe containing only dust and radiation is given by
\begin{equation}
a(\eta)=a_{\rm{eq}}\eta(\eta+2),
\end{equation}
where $a_{\rm{eq}}$ is the scale factor at equality.

To move towards the standard WKB form, we make the transformation $h=y/[\eta(\eta+2)]$ which eliminates the first derivative term leading to
\begin{equation}
\ddot{y}+\left(q^2-\frac{\ddot{a}}{a}\right)=0.
\end{equation}
Before proceeding, we notice that this transformation is singular at $\eta=0$, the point at which our boundary conditions are specified.  This prevents us from applying the boundary conditions (b.c.) directly and will motivate looking for asymptotic solutions to \eqref{heom_ap} which we will discuss later.

We next map from the interval $[0,\infty]$ to $[-\infty,\infty]$ by the transformations $\eta=\exp(x)$ and $u(x)=\exp(-x/2)y(x)$.  These place our problem in the WKB form
\begin{equation}\label{wkbform}
u''(x)=f(x)u(x),
\end{equation}
\begin{equation}\label{wkbfunc}
f(x)=\frac{1}{4}+\frac{2e^x}{e^x+2}-q^2e^{2x}.
\end{equation}

The standard WKB problem \cite{Bender:1978}
\begin{equation}\label{wkbstandard}
\epsilon^2 y''(x)=Q(x)y(x),
\end{equation}
where $Q(0)=0$, $Q(x)>0$ for $x<0$ and $Q(x)<0$ for $x>0$,
has the uniform Langer solution
\begin{multline}\label{wkblanger}
y(x)= \sqrt{\pi}\,\left(\frac{3}{2\epsilon}S_0\right)^{1/6}[Q(x)]^{-1/4}\\ \times\left\{2C_2\rm{Ai}\left[\left(\frac{3}{2\epsilon}S_0\right)^{2/3}\right]+
C_1 \rm{Bi}\left[\left(\frac{3}{2\epsilon}S_0\right)^{2/3}\right]\right\},
\end{multline}
with
\begin{equation}
S_0=\int_x^0\sqrt{Q(t)} dt.
\end{equation}
Here, $\rm{Ai}(x)$ and $\rm{Bi}(x)$ are Airy functions and $C_1$ and $C_2$ are constant coefficients to be set by the boundary conditions.

Applying this directly to Eqs. \eqref{wkbform} and \eqref{wkbfunc} and manipulating the algebra slightly, we obtain
\begin{multline}\label{hwkb_ap}
h_{\rm{wkb}}(\tau)=\frac{\sqrt{\pi}\,\Gamma(k\tau)^{-1/4}}{\tau^{1/2}(\tau+2)}\left(\frac{3}{2}S_0(\tau)\right)^{1/6} \\ \times\left\{2C_2 \rm{Ai}\left[\left(\frac{3}{2}S_0(\tau)\right)^{2/3}\right]+C_1 \rm{Bi}\left[\left(\frac{3}{2}S_0(\tau)\right)^{2/3}\right] \right\},
\end{multline}
with
\begin{equation}
\Gamma(s)=\frac{1}{4}+\frac{2s}{s+2k}-s^2,
\end{equation}
and
\begin{equation}
S_0(\tau)=\int_{k\tau}^{k\tau_T} \sqrt{\Gamma(s)}\, \frac{ds}{s}.
\end{equation}
Here, $\tau_T$ is the solution to $\Gamma(k\tau)=0$.  It can be shown that $k\tau_T$ is bounded such that $1/2\le k\tau_T \le 3/2$.  These expressions may be evaluated directly when $\tau<\tau_T$, but some care must be taken when $\tau>\tau_T$. In this case, $\Gamma(k\tau)<0$, and we must keep careful track of minus signs.  Making the appropriate manipulations, we obtain for $\tau>\tau_T$
\begin{multline}
h_{\rm{wkb}}(\tau)= \frac{\sqrt{\pi}\,[-\Gamma(k\tau)]^{-1/4}}{\tau^{1/2}(\tau+2)}\left(\frac{3}{2}S_0(\tau)\right)^{1/6} \\ \times \left\{2C_2 \rm{Ai}\left[-\left(\frac{3}{2}S_0(\tau)\right)^{2/3}\right]+C_1 \rm{Bi}\left[-\left(\frac{3}{2}S_0(\tau)\right)^{2/3}\right] \right\},
\end{multline}
with
\begin{equation}
\Gamma(s)=\frac{1}{4}+\frac{2s}{s+2k}-s^2,
\end{equation}
and
\begin{equation}
S_0(\tau)=\int_{k\tau_T}^{k\tau} \sqrt{-\Gamma(s)}\, \frac{ds}{s}.
\end{equation}

Asymptotic expansions of the Airy functions recover the more familiar exponential and trigonometric forms for the WKB connection formula.

Having obtained an expression for $h$, we now need to apply the boundary conditions.  Unfortunately, if we naively try to apply the boundary conditions, we discover that $\dot{h}_{\rm{wkb}}(\tau=0)$ is divergent.  This is a consequence of the transformation required to place the equation of motion into WKB form which is singular at $\tau=0$.  To get around this problem, we seek to apply the boundary conditions at some small time when the gravitational-wave has had little chance to evolve and will be well described by a series solution.  

Eq. \eqref{heom_ap} has a regular singular point at $\tau=0$ so we attempt a Frobenius series solution of the form 
\begin{equation}
h(\tau)=\tau^\sigma\sum_{n=0}^{\infty} a_n \tau^n.
\end{equation}
Substituting this expression into Eq. \eqref{heom_ap}, solving the indicial equation for $\sigma$, and equating like powers of $\tau$ to get a recurrence relation leads to an asymptotic polynomial expression for $h$.  In fact, this procedure only generates one of the two linearly independent solutions to Eq. \eqref{heom_ap}.  Application of the boundary conditions causes the other solution to vanish and normalises this one leaving us with a solution,
\begin{equation}
h(\tau)=1-\frac{q^2\tau^2}{6}+\frac{q^2\tau^3}{36}-\frac{q^2}{240}(3-2q^2)\tau^4.
\end{equation}
This is valid only when $\tau\ll1$.  We can use this to extrapolate the b.c. from zero time to some small time and use it to determine the constant coefficients in Eq. \eqref{hwkb_ap}.  

\begin{figure}[htbp]
\begin{center}
\includegraphics[scale=0.4]{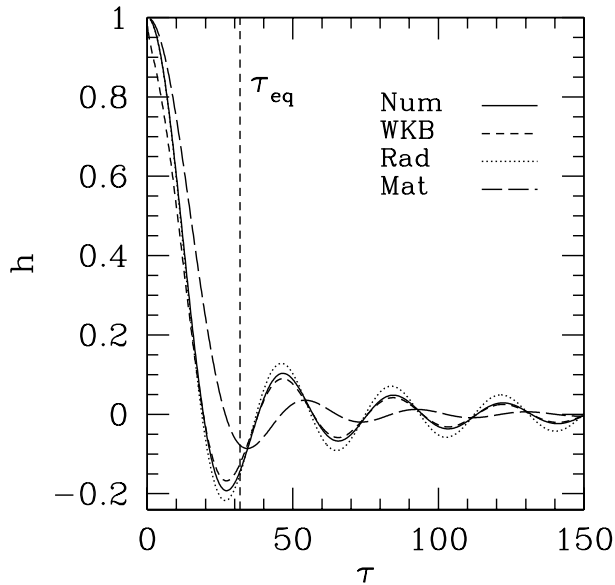}
\caption{Comparison of the WKB (dashed curve) and numerical (solid curve) evolution of $h$ for $k=0.1677\, \rm{Mpc}^{-1}$ in the fiducial cosmology.  The amplitudes $h_{\rm{mat}}$ (long dashed curve) and $h_{\rm{rad}}$ (dotted curve) are plotted for reference, and $\tau_{\rm{eq}}$ is indicated by a vertical line.}
\label{fig:wkbplot}
\end{center}
\end{figure}
The WKB solution (Fig. \ref{fig:wkbplot}) accurately reproduces the phase of the oscillation in the regime  $\tau>\tau_T$ although it underestimates the amplitude of $h$ in this region by a factor of $\sim0.87$.  The approximation is also good for $\tau<\tau_T$ failing only at times comparable to the time at which the b.c. are applied.  For the power spectra calculated in this paper, this is taken to be $k\tau_{\rm{match}}=10^{-5}$.  Consequently, the power spectrum become unreliable at small multipoles, $l<10$.  On these scales power is generated by modes that have evolved very little by the time of recombination.




\end{document}